\documentstyle[epsfig,12pt]{procsla}
\def\beq{\begin{eqnarray}}
\def\eeq{\end{eqnarray}}
\newcommand{\lsim}{\raisebox{-4pt}{$\,\stackrel{\textstyle <}{\sim}\,$}}

 \def\@makefnmark{\hbox to 0pt{$^{\@thefnmark}$\hss}}

\begin{document}

\begin{flushright}
                                                      hep-ph/9710385\\ 
                                                          WU-B 97/29\\
\end{flushright}
\vskip-3em

{\Large\bf The $\eta_c\gamma$ Transition Form Factor\footnote
 {Contribution to {\it IV$^{th}$ International Workshop on  Progress
 in Heavy Quark Physics}\/,\\ Rostock,~September~20-22,~1997.}}\\[2mm]
{\large\it Thorsten~Feldmann\footnote
 {Supported  by {\it Deutsche Forschungsgemeinschaft}}}\\[2mm] 
{\small Department of Theoretical Physics, University of Wuppertal,\\ 
  D-42097 Wuppertal, Germany}\\[2mm]

\begin{abstract}
The $\eta_c\gamma$ transition form factor is calculated within a
perturbative approach.  It is shown that the $Q^2$ dependence of the
form factor is well determined in the region where experimental data
is expected in the near future.
\end{abstract}

\normalsize

\section{Introduction}

\begin{figure}[hbt]
\begin{center}
\epsfclipon
\parbox[c]{0.44\textwidth}
{\psfig{file=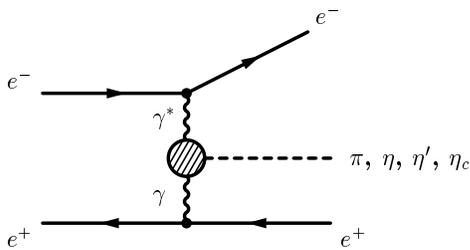, bb=200 590 470 740, width=0.4\textwidth}\hfill}
\hskip2em
\epsfclipoff
\parbox[c]{0.44\textwidth}
{\psfig{file=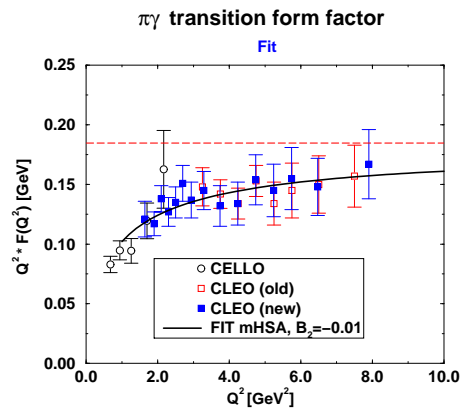, width=0.333\textwidth, angle=-90}\hfill}
\end{center}
\caption{
Meson-Photon transition form factors in $e^+e^-$ collisions (left).
The $\pi\gamma$ form factor: experimental data and mHSA fit (right).}
\label{Profig}
\end{figure}

Pseudoscalar meson-photon transition form factors (see
Fig.~\ref{Profig}) at large momentum transfer $Q^2$ have attracted the
interest of many theoreticians during the last years, stimulated by
the CLEO measurements\cite{CLEO95,CLEO97}.  At the upper end of the
measured $Q^2$ range  the CLEO data for the $\pi\gamma$ form factor
only deviate by about 15\% from the limiting value $\sqrt2f_\pi/Q^2$
which is predicted by QCD\cite{BrLe80}.  The data allow for a rather
precise determination of the pion's light-cone wave
function\cite{KrRa96,JaKrRa96,MuRa97}; and we find, within the
modified hard scattering approach\cite{KrRa96,bot:89} (mHSA), a value
of $-0.01\pm0.1$ for the expansion coefficient $B_2$ at the scale
$\mu=1$~GeV (see Fig.~\ref{Profig}).  The situation is more
complicated for  the $\eta\gamma$ and the $\eta'\gamma$ form factors
due to the mixing and $SU(3)_F$ flavor symmetry breaking.  A
determination of the decay constants and the mixing angle from the
$\eta\gamma$ and $\eta'\gamma$ transition form factors is  also
possible\cite{JaKrRa96,FeKr97b}.

There is a fourth form factor of the same type, namely the
$\eta_c\gamma$ form factor which is neither experimentally nor
theoretically known. Since a measurement of that form factor up to a
momentum transfer of about 10~GeV$^2$ seems feasible\cite{Au96},  a
theoretical analysis and prediction of it is desirable and has been
performed by us recently\cite{FeKr97a}.

\section{The perturbative approach}
\label{sec:2}

In analogy to the $\pi\gamma$ case\cite{KrRa96,JaKrRa96} we employ a
perturbative approach on the basis of a factorization of short- and
long-distance physics\cite{BrLe80}.  Observables are then described as
convolutions of a perturbatively calculable hard scattering amplitude
$T_H$ and a universal (process-independent) hadronic light-cone wave
function $\Psi$ of the $\eta_c$'s leading $c\bar c$ Fock
state\footnote{Higher Fock state contributions to the $\eta_c\gamma$
form factor are suppressed by powers of $\alpha_s/m_c^2$. However, higher
Fock states can be important in other decays of heavy
quarkonia\cite{BoKrSc97a,BoKrSc97b,Br97}.},  which embodies soft
non-perturbative physics,
\beq 
F_{\eta_c\gamma} (Q^2) &=&  
\int_0^1 dx \, \int \frac{d^2 {\vec k_\perp}}{16 \pi^3}  
\, \Psi(x, {\vec k_\perp}) \, T_H(x,{\vec k_\perp},Q) \ .
\label{Fdef2}
\eeq
Here $x$ is the usual meson's momentum fraction carried by the $c$
quark, and $\vec k_\perp$ denotes its transverse momentum.  In the
present case the mass of the charm quarks already provides a large
scale, which allows the application of the perturbative approach even
for zero virtuality of the probing photon, $Q^2\to 0$; and  for heavy
quarks a Sudakov factor in (\ref{Fdef2}) can be ignored\cite{FeKr97b}.
The hard scattering amplitude in leading order is  easily calculated.
With one photon being almost on-shell ($q_1^2 \simeq 0$) and the
virtuality of the second photon denoted as $q_2^2 = -Q^2$, this leads
to (with $\bar x = (1-x)$)
\beq 
T_H(x,{\vec k_\perp},Q) &=&  
  \frac{e_c^2 \, 2 \sqrt 6}
    {x\, Q^2 +m_c^2 + x \, \bar x  \, M_{\eta_c}^2 + {\vec k_\perp^2}}
  + (x \leftrightarrow \bar x) + O(\alpha_s)
\label{THhat2}
\eeq
where $M_{\eta_c}$ ($=2.98~{\rm GeV}$) is the mass of the $\eta_c$
meson,  and $m_c \simeq M_{\eta_c}/2$ is the charm quark mass.  The
charge of the charm quark in units of the elementary charge is denoted
by $e_c$.  For the $\eta_c$  wave function,
\beq 
\Psi(x, {\vec k_\perp}) = 
  \frac{f_{\eta_c}}{2 \sqrt{6}} \,
  \phi(x) \, \Sigma({\vec k_\perp}) \ ,
\label{psidef}
\eeq
we use a form adapted from Bauer, Stech and Wirbel\cite{BSW85}.  Here
$f_{\eta_c}$ is the decay constant (corresponding to $f_\pi =
131$~MeV), and $\phi(x)$ is the quark distribution amplitude which is
parameterized as
\beq 
\phi(x) &=& 
N_\phi \ x \, \bar x \ 
\exp\left[ - a^2 \, M_{\eta_c}^2 \, \left(x - x_0\right)^2\right] \ .
\label{da}
\eeq
The normalization constant $N_\phi$ is determined from the usual
requirement $\int_0^1 dx \, \phi(x)~=~1$.  The distribution amplitude
(\ref{da}) exhibits a pronounced maximum at $x_0$ and is exponentially
damped in the endpoint regions.  Furthermore, $\Sigma$ is a Gaussian
shape function which takes into account the finite transverse size of
the meson,
\beq 
\Sigma({\vec k_\perp})  &=& 
16 \pi^2 \, a^2 \, \exp[-a^2 \, {\vec k_\perp^2}] \ .
\label{Sigmafunc}
\eeq
The decay constant of the $\eta_c$ meson is not accessible  in a
model-independent way at present.  Usually, one estimates $f_{\eta_c}$
in a non-relativistic approach which provides a connection between
$f_{\eta_c}$ and the well-determined decay constant of the $J/\psi$,
$f_{\eta_c} \simeq f_{J/\psi} = 409$~MeV.  However, the  $\alpha_s$
corrections  are large\cite{Ba79},  and the relativistic corrections
are usually large and model-dependent.

The parameters entering the wave function are further  constrained by
the Fock state probability $P_{c\bar c}$.  One expects $0.8 \leq
P_{c\bar c} < 1$ for a charmonium state (for smaller values of
$P_{c\bar c}$ one would not understand the success of non-relativistic
potential models for these states).  Since the perturbative
contribution to the $\eta_c\gamma$ form factor only mildly depends on
the value of $P_{c\bar c}$, we use $P_{c\bar c}=0.8$ as a constraint
for the transverse size parameter $a$.  For $f_{\eta_c}=409$~MeV this
leads  to a reasonable value\cite{FeKr97b}, $a = 0.97$~GeV$^{-1}$.

The  two photon decay width $\Gamma[\eta_c\to\gamma\gamma]$, the
experimental value of which still suffers from large
uncertainties\cite{PDG96}, can be directly related to the
$\eta_c\gamma$ transition  form factor at $Q^2=0$
\beq 
\Gamma[\eta_c \to \gamma\gamma] &=& 
\frac{\pi \alpha^2 M_{\eta_c}^3}{4}  \left| F_{\eta_c\gamma}(0) \right|^2  
\label{eq:etacwidth}
\eeq
One may use this decay rate as a normalization condition for
$F_{\eta_c\gamma}(Q^2=0)$ and present the result in the form
$F_{\eta_c\gamma}(Q^2)/F_{\eta_c\gamma}(0)$.  In this way the
perturbative QCD corrections at $Q^2=0$ to the $\eta_c\gamma$
transition form factor are automatically included, and also the
uncertainties in the present knowledge of $f_{\eta_c}$ do not enter
our predictions.

\section{Results and Conclusions}
\label{sec:4}

\begin{figure}[hbtp]
\unitlength1cm
\begin{center}
\epsfclipoff 
\parbox[c]{0.44\textwidth} 
{\psfig{file=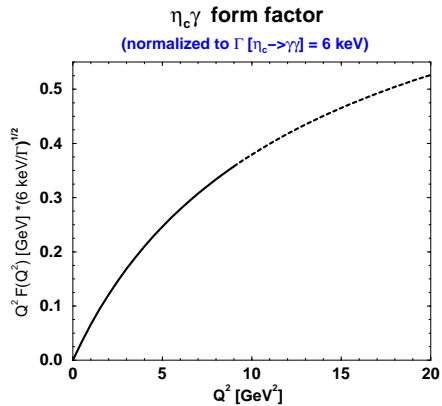, width=0.333\textwidth, angle = -90}}
\end{center}
\caption{
The predictions for   $Q^2 \, F_{\eta_c\gamma}(Q^2)$  scaled to
 $\Gamma[\eta_c\to\gamma\gamma] = 6$~keV in the leading order of the
 perturbative approach (for $P_{q\bar q}$=0.8).  The dashes indicate
 the $Q^2$ region where QCD corrections may alter the predictions
 slightly.}
\label{fig}
\end{figure}
In Fig.~\ref{fig} we present the result for the transition form factor
$Q^2 \, F_{\eta_c\gamma}$ scaled to a partial width
$\Gamma[\eta_c\to\gamma\gamma]$ of $6$~keV.  In order to discuss the
qualitative features of our result in a rather simple fashion, one can
restrict oneself to first order corrections to the collinear ($\vec
k_\perp^2 \simeq 0$) and peaking approximation ($x\simeq x_0$), which
can be expressed by the small quantity $\langle
k_\perp^2\rangle=1/2a^2 \ll M_{\eta_c}^2$.  For $Q^2 \leq
M_{\eta_c}^2$ one then obtains the following
approximation\cite{FeKr97a}
\beq 
F_{\eta_c\gamma}(Q^2) &\simeq& 
\frac{4\,e_c^2 \, f_{\eta_c} }
 {Q^2 + M_{\eta_c}^2 + 2 \, \langle \vec k_\perp^2 \rangle } 
\simeq
 \frac{F_{\eta_c\gamma}(0)} 
{1 + Q^2/(M_{\eta_c}^2 + 2\, \langle \vec k_\perp^2 \rangle)} \ ,
\label{compact2}
\eeq
which reveals that, to a very good approximation, the predictions for
the scaled $\eta_c\gamma$ form factor are rather insensitive to the
details of the wave function. Only the mean transverse momentum
following from it is required, leading to an effective pole mass of
$\sqrt{M_{\eta_c}^2 + 2 \langle k_\perp^2\rangle}= 3.15~{\rm GeV}$
which is very close to the  value of the $J/\psi$ mass that one would
have inserted in the vector meson dominance model.  The deviation from
the full result amounts only to 4\%  at $Q^2=10$~GeV$^2$,  which is
likely smaller than the expected experimental errors in a future
measurement of the $\eta_c\gamma$ form factor\cite{Au96}.  These
considerations nicely illustrate that the $Q^2$ dependence of the
$\eta_c\gamma$ form factor is well determined.  The main uncertainty
of the prediction resides in the normalization, i.e.\ the $\eta_c$
decay constant or the value of the form factor at $Q^2=0$.

Let us briefly discuss, how $\alpha_s$ corrections  may modify the
leading order result for the $\eta_c\gamma$ form factor: One has to
consider two distinct kinematic regions.  First, if $Q^2 \lsim
M_{\eta_c}^2$ one can neglect the evolution of the wave function, and
one is left with the QCD corrections to the hard scattering amplitude
$T_H$, which have been calculated\cite{ShVy81} in the peaking and
collinear approximation to order $\alpha_s$.  For the scaled form
factor the $\alpha_s$ corrections at $Q^2$ and at $Q^2=0$ cancel to a
high degree, and even at $Q^2=10$~GeV$^2$ the effect of the $\alpha_s$
corrections is less than 5\%.

Secondly, for $Q^2 \gg M_{\eta_c}^2$, one can neglect the quark and
meson masses and arrives at the same situation as for  the pions. The
$\alpha_s$ corrections to the hard  scattering amplitude and the
evolution of the wave function with $Q^2$ are
known\cite{BrLe80,ShVy81,Br83}.  For very large values of $Q^2$ the
asymptotic behavior of the transition form factor is completely
determined by QCD, since any meson distribution amplitude evolves into
the asymptotic form $\phi(x)\to\phi_{\rm as}(x)=6 \, x \, \bar x$, 
\beq 
F_{\eta_c\gamma}(Q^2) \to  
\frac{2 e_c^2 f_{\eta_c}}{Q^2}
\, \int_0^1 dx \, \frac{\phi(x)}{x} 
\to \frac{8 \, f_{\eta_c}}{3 \, Q^2} \ . 
  \qquad (\ln Q^2 \to \infty)
\label{asymp}
\eeq

A precise measurement of the strength of the $\eta_c\gamma$ transition
form factor may serve to determine the decay constant $f_{\eta_c}$
(see  (\ref{compact2})). Though attention must be paid to the fact
that the obtained value of $f_{\eta_c}$ is subject to large QCD
corrections (about of the order 10-15\% for $Q^2\lsim 10$~GeV$^2$)
which should be taken into account for an accurate extraction of the
$\eta_c$ decay constant.

\bibliography{kroll,ref}

\end{document}